\documentclass{rcdj}

\usepackage{hhline}

\renewcommand{\theparagraph}{\arabic{paragraph}}

\begin{document}

\setcounter{page}{21}

\title{GENERALIZATION OF THE\\ GORYACHEV\f CHAPLYGIN CASE}
\runningtitle{GENERALIZATION OF THE GORYACHEV\f CHAPLYGIN CASE}
\runningauthor{A.\,V.\,BORISOV, I.\,S.\,MAMAEV}
\authors{A.\,V.\,BORISOV}
{Department of Theoretical Mechanics\\
Moscow State University,
Vorob'ievy Gory\\
119899, Moscow, Russia\\
E-mail: borisov@rcd.ru}
\authors{I.\,S.\,MAMAEV}
{Laboratory of Dynamical Chaos and Nonlinearity\\
Udmurt State University, Universitetskaya, 1\\
426034, Izhevsk, Russia\\
E-mail: mamaev@rcd.ru}
\abstract{In this paper we present a generalization of the Goryachev\f
Chaplygin integrable case on a bundle of Poisson brackets, and on Sokolov
terms in his new integrable case of Kirchhoff equations. We also present a
new analogous integrable case for the quaternion form of rigid body
dynamics' equations. This form of equations is recently developed and we
can use it for the description of rigid body motions in specific force
fields, and for the study of different problems of quantum mechanics. In
addition we present new invariant relations in the considered problems.}
\journal{REGULAR AND CHAOTIC DYNAMICS, V.\,7, \No1, 2002}
\received 20.12.2001.
\amsmsc{37J35, 70E17}
\doi{10.1070/RD2002v007n01ABEH000192}
\maketitle

In this paper we present a generalization of the Goryachev\f Chaplygin
integrable case on a bundle of Poisson brackets including (co)algebras
$so(4)$ and~$so(3,1)$, and on the quaternion form of Euler\f Poisson
equations. Note that the generalization on the bundle is connected with
the introduction of variables on the bundle analogous to Andoyer\f Deprit
variables. These variables are separating for all members of the bundle.
We also obtain the $L-A$-pair of the generalized Goryachev\f Chaplygin
case of quaternion equations. In a particular case this pair is reduced to
the $L-A$ pair of the classical Goryachev\f Chaplygin case.

\paragraph{}\label{punt1}
Let us consider a generalization of the classical Goryachev\f Chaplygin
case of Euler\f Poisson equations
$$
\dot {\bM}=\bM\times \pt{H}{\bM}+\bs\gamma\times\pt{H}{\bs\gamma},\quad
\dot{\bs\gamma}=\bs\gamma\times\pt{H}{\bM}
$$
for the zero value of the area integral~$(\bM,\bs\gamma)=0$. We add a
gyrostatic moment and singular term to the Hamiltonian
\begin{equation}
\label{gen5_1}
H=\frac12(M_1^2+M_2^2+4M_3^2)+\lambda M_3+\mu \gamma_1+\frac12\frac{a}
{\gamma_3^2},
\end{equation}
where~$\lambda,\mu,a=\const$. The additional integral is of order three
with respect to the moments
$$
F=\Bigl(M_3+\frac{\lambda}{2}\Bigr)\Bigl(M_1^2+M_2^2+\frac{a}{\gamma_3^2}\Bigr)
-\mu M_1\gamma_3.
$$
D.\,N.\,Goryachev himself in the paper~\cite{goriachev1} showed the
generalization~\eqref{gen5_1} on zero value surface of gyrostatic
moment~$\lambda=0$ (for $\lambda\ne 0$, $a=0$ it was shown by
L.\,N.\,Sretensky~\cite{sret}). The complete
form of generalization~\eqref{gen5_1} was considered by
I.\,V.\,Komarov and
V.\,B.\,Kuznetsov~\cite{KomKuz}. They also
presented some quantum mechanical interpretation of the singular term.

The Poisson bracket of variables $\bM,\bs\gamma$ is defined by algebra
$e(3)$ and has the following form
\begin{equation}
\label{eq-x1}
\{M_i,M_j\}=-\eps_{ijk}M_k,\quad \{M_i,\gamma_j\}=-\eps_{ijk}\gamma_k,
\quad \{\gamma_i,\gamma_j\}=0.
\end{equation}

\paragraph{}\label{punt2}
Let us present the generalization of the Goryachev\f Chaplygin case on the
bundle of brackets of the form
\begin{equation}
\label{eq-x2}
\{M_i,M_j\}=-\eps_{ijk}M_k,\quad \{M_i,\gamma_j\}=-\eps_{ijk}\gamma_k,
\quad \{\gamma_i,\gamma_j\}=-x\eps_{ijk}M_k.
\end{equation}
At $x=1$ these commutation relations correspond to the algebra $so(4)$.

We can present the Casimir functions of bracket~\eqref{eq-x2} in the form
\begin{equation}
\label{eq-x2-1}
F_1=(\bM,\bs\gamma),\quad F_2= \bs\gamma^2+x\bM^2.
\end{equation}

Now let us construct the variables on bracket~\eqref{eq-x2} analogous to
the Andoyer\f Deprit variables~\cite{bormam2}.

Suppose the component~$M_3$ is equal to the momentum
\eq[m6.1eul]
{
L=M_3.
}
The variable~$l$ canonically conjugate to the variable~$L$ $(\{l,\,L\}=1)$
on subalgebra~$so(3)$ with the generators~$M_1,M_2,M_3$ is constructed by
integration of the Hamiltonian flow with Hamiltonian function~$\cH=L$
\eq[m6.2eul]{
\arr{
 \frac{dM_1}{dl} =\{M_1,\,L\}=M_2,\quad \frac{dM_2}{dl}  =\{M_2,\,L\}=-M_1,\\
 \frac{dM_3}{dl} =\{M_3,\,L\}=0.
}
}
Hence using the commutation relation~$\{M_1,\,M_2\}=-M_3$ we obtain
\eq[m6.3eul]{
M_1=\sqrt{G^2-L^2}\sin l,\qq M_2=\sqrt{G^2-L^2}\cos l,
}
where~$G^2=M_1^2+M_2^2+M_3^2$ is the Casimir function of
subalgebra~$so(3)$.

Suppose~$G$ is the second momentum and construct the canonically conjugate
variable~$g$. We chose~$H=G$ as a new Hamiltonian, and the corresponding
flow on the whole bundle $\cL_x$ has the form
\eq[m6.4eul]{
\frac{d\bs M}{dg}=0,\qq \frac{d\bs \gamma}{dg}=\frac 1G \bs \gamma\times \bs M,
}
where~$g$ is the variable canonically conjugate to~$G$.

According to~(\ref{m6.4eul}),~$\bs M$ does not depend on~$g$, and using
equations~(\ref{m6.4eul}) and Casimir functions~(\ref{eq-x2-1}) we obtain
for~$\bs\gamma$ the equations
\eq[m6.5eul]{\arr{
\bs\gamma=\frac{H}{G^2}\bs M+\frac{\al}{G}(\bs M\times \bs e_3\sin g+G\bs
M\times(\bs M\times \bs e_3)\cos g),\\
\al^2=\frac{c_2-xG^2-\frac{H^2}{G^2}}{G^2-L^2},\qq \bs e_3=(0,\,0,\,1),
}}
where $c_2=x\bM^2+\bs\gamma^2$, and $H$ is the traditional notation of the
area integral~$H=(\bM,\bs\gamma)=c_1$.

Thus~(\ref{m6.1eul}),~(\ref{m6.3eul}),~(\ref{m6.5eul}) define the
symplectic variables on the whole bundle~$\cL_x$ corresponding at~$x=0$,
$c=1$ to the well-known Andoyer\f Deprit variables in rigid body dynamics.

\paragraph{}\label{punt3}
Using~(\ref{m6.1eul}), (\ref{m6.5eul}) we obtain the generalization of the
particular Goryachev\f Chaplygin integrable case on the bundle~$\cL_x$.
We chose the Hamiltonian in the form
\eq[m6.7eul]{
\cH=\frac 12(G^2+3L^2)+\lm L+a(\cos l\cos g+\frac LG\sin l\sin g),
}
where~$a,\,\lm$ are constants.

In~(\ref{m6.7eul}) we add the linear with respect to~$L$ term. It is
interpreted on algebra $e(3)$ as the component of the gyrostatic
moment~\cite{sret}.

{\itshape We can separate variables in system~\eqref{m6.7eul}.} Indeed,
let us apply the canonical change of variables
\begin{equation}
\label{m6.g4}
 L=p_1+p_2, \q G=p_1-p_2, \qq q_1=l+g,\q q_2=l-g.
\end{equation}
Now Hamiltonian~(\ref{m6.7eul}) can be presented in the form
\eq[m6.8eul]{
\cH=\frac 12\frac{p_1^3-p_2^3}{p_1-p_2}-\lm\frac{p_1^2-p_2^2}{p_1-p_2}+
\frac{a}{p_1-p_2}(p_1\sin q_1+p_2\sin q_2).
}
Using \eqref{m6.1eul}, \eqref{m6.3eul}, \eqref{m6.5eul} we present
Hamiltonian~(\ref{m6.7eul}) as a function of variables~$\bs M,\,\bs
\gamma$ on the zero surface of area integral~$(\bs M,\,\bs \gamma)=H=0$
\eq[gen5_7]
{
 \cH=\frac 12(M_1^2+M_2^2+4M_3^2)+\lm M_3+\mu\frac{\gamma_1}{|\bs \gamma|}.
}

The additional integral in this case has the form
\begin{equation}
\label{gen.m7.17}
 F=\Bigl(M_3+\frac{\lm}{2}\Bigr)(M_1^2+M_2^2)-\mu M_1\frac{\gamma_3}
{|\bs \gamma|}.
\end{equation}

On algebra~$e(3)$ we have $|\bs \gamma|=1$ and obtain the classical
Goryachev\f Chaplygin integrable case. In the classical case this method
of variables' separation was suggested by
{\itshape V.\,B.\,Kozlov}~\cite{KozlovMethods}.

Singular term~\eqref{gen5_1} is generalized on the bundle in the following
way:
$$
\begin{gathered}
H= \frac12(M^2_1+M_2^2+4M_3^2)+\lambda M_3 +\mu
\frac{\gamma_1}{|\bs\gamma|}+\frac12\frac{a\bs\gamma^2}{\gamma_3^2},\\
F=\Bigl(M_3+\frac{\lambda}{2}\Bigr)\Bigl(M_1^2+M_2^2+a\frac{\gamma_2}{\gamma_3^2}\Bigr)-
\mu M_1\gamma_3,
\end{gathered}
$$
although variables' change~\eqref{m6.g4} in this case doesn't produce the
separation of variables (at least we don't know such separation).

\paragraph{}\label{punt4}
Recently Sokolov and Tsyganov present a new generalization of particular
integrable case~\eqref{gen5_1} on bracket~\eqref{eq-x1} with the
Hamiltonian containing the quadratic cross terms with respect
to~$\bM,\bs\gamma$.

The most general form of the integrable family in this case is presented
as the following Hamiltonian
\begin{equation}
\label{eq4-*1}
\begin{gathered}
H=\frac12\Bigl(M_1^2+M_2^2+4M_3^2+\frac{\eps}{\gamma_3^2}\Bigr)+\lambda
M_3+\mu_1\gamma_1+\mu_2\gamma_2+\\
+a_1(2M_3\gamma_1-M_1\gamma_3)+a_2(2M_3\gamma_2-M_2\gamma_3).
\end{gathered}
\end{equation}
The additional integral is
\begin{equation}
\label{eq4-*2}
F=\Bigl(M_3+a_1\gamma_1+a_2\gamma_2+\frac{\lambda}{2}\Bigr)\Bigl(
M_1^2+M_2^2+\frac{\eps}{\gamma_3^2}\Bigr)-(\mu_1 M_1+\mu_2M_2)\gamma_3.
\end{equation}
At $\eps=0$ we can separate variables in Hamiltonian~\eqref{eq4-*1} using
the linear combination of Andoyer\f Deprit variables analogous
to~\eqref{m6.g4}. Indeed, we can show that
\begin{equation}
\label{eq4-*3}
\begin{gathered}
H=\frac{1}{p_1-p_2}\bigl(f(p_1,q_1)-g(p_2,q_2)\bigr),\\
f=2a_1p_1^3+2\Bigl(\frac{\lambda}{2}+a_1\sin q_1+a_2\cos q_2\Bigr)p_1^2+
(\mu_1\sin q_1+\mu_2\cos q_1)p_1,\\
g=2a_1p_2^3+2\Bigl(\frac{\lambda}{2}-a_1\sin q_2-a_2\cos q_2\Bigr)p_2^2-
(\mu_1\sin q_2+\mu_2\cos q_2)p_2.
\end{gathered}
\end{equation}
As we note above, we don't know such separation at $\eps\ne 0$.

If the gravity field is absent $(\mu_1=\mu_2=0)$, then the integral is
presented as a product of factors $F=k_1k_2$, where
\begin{equation}
\label{eq4-*6}
k_1=a_0 M_3+a_1\gamma_1+a_2\gamma_2+\frac{\lambda}{2},\quad
k_2=M_1^2+M_2^2+\frac{\eps}{\gamma_3^2}.
\end{equation}
We can easily show that the equations of motion for $k_1$, $k_2$ have the
form
$$
\dot k_1=2(a_1\gamma_2-a_2\gamma_1)k_1,\quad \dot
k_2\Bigr|_{(\bM,\bs\gamma)=0}=-2(a_1\gamma_2-a_2\gamma_1)k_2.
$$
Thus the equations $k_1=0$ and $k_2=0$ define the invariant relations of
system~\eqref{eq4-*1} at $\mu_1=\mu_2=0$ (and $k_2=0$ is the particular
invariant relation on zero level of~$(\bM,\bs\gamma)=0$, and $k_1=0$ is
general invariant relation). It seems that, invariant
relations~\eqref{eq4-*6} and the additional integral in the
form~$F=k_1k_2$ where introduced by authors.

\paragraph{}
Using the observation from the first section we can generalize integrable
case~\eqref{eq4-*1} and invariant relations on bundle of
brackets~\eqref{eq-x2}. We shall make the following substitution
in Hamiltonian~\eqref{eq4-*1}, integral~\eqref{eq4-*2}, and  invariant
relations~\eqref{eq4-*6}
$$
\gamma_i\to \frac{\gamma_i}{|\bs\gamma|},\quad |\bs\gamma|=\sqrt{\gamma_1^2+\gamma_2^2+
\gamma_3^2}.
$$
In this case at $\eps=0$ we can also separate variables using the analog
of Andoyer\f Deprit variables and substitution~\eqref{m6.g4}.

We should note that unlike the generalization of Kovalevskaya
case~\cite{bormam2}, where the Hamiltonian and integral explicitly depend
on the parameter of bundle~$x$, in the generalization of Goryachev\f
Chaplygin case the integrals don't depend on the parameter of bundle.

\paragraph{}
Let us consider another possibly not so natural case of motion equations
of rigid body with a potential linear with respect to Rodrig\f Hamilton
parameters and not with respect to direction cosines~\cite{bormam2}
\begin{equation}
\label{kvat-eq-1}
H=\frac12({\bf A}\bM,\bM)+\sum_{i=0}^3r_i\lambda_i,\quad r_i=\const,
\end{equation}
The equations of motion in this case are
\begin{equation}
\label{eq-d1}
\begin{gathered}
\dot{\bM}=\bM\times \pt{H}{\bM}+\frac12\bs\lambda\times\pt{H}{\bs\lambda}+
\frac{1}{2}\pt{H}{\lambda_0}\bs\lambda-\frac{1}{2}\lambda_0\pt{H}{\bs\lambda},\\
\dot\lambda_0=-\frac{1}{2}\Bigl(\bs\lambda,\pt{H}{\bM}\Bigr),\quad
\dot{\bs\lambda}=\frac12
\bs\lambda\times\pt{H}{\bM}+\frac12\lambda_0\pt{H}{\bM},
\end{gathered}
\end{equation}
where $\bs\lambda=(\lambda_1,\lambda_2,\lambda_3)$.

These equations are Hamiltonian one with the Poisson bracket
\begin{equation}
\label{eq-d2}
\begin{gathered}
\{M_i,M_j\}=-\eps_{ijk}M_k,\quad \{M_i,\lambda_0\}=\frac12\lambda_i,\\[10pt]
\{M_i,\lambda_i\}=-\frac12(\eps_{ijk}\lambda_k+\delta_{ij}\lambda_0),\quad
i,j,k=1,2,3,
\end{gathered}
\end{equation}
corresponding to the Lie algebra $so(3)\otimes_s\mR^4$. The relations
between quaternions and Euler angles have the form
$$
\begin{aligned}
\lambda_0 & =\cos\frac{\theta}{2}\cos\frac{\psi+\vfi}{2},\quad &
\lambda_1 & =\sin \frac{\theta}{2}\cos\frac{\psi-\vfi}{2},\\[10pt]
\lambda_2 & =\sin\frac{\theta}{2}\sin\frac{\psi-\vfi}{2},\quad &
\lambda_3 & =\cos \frac{\theta}{2}\sin\frac{\psi+\vfi}{2}.
\end{aligned}
$$
In the classical mechanics such potential aren't presented because its
dependance on the body position is nonunique (more exactly it is
double-valued function). To justify the study of such equations we
can refer to the problems of quantum mechanics and point masses dynamics in
curved space~$S^3$, and on some formal technics of ${\bf L-A}$\1pair
construction~\cite{bormam2,BorisovMamaev}. It turns out that after the
reduction of order of system~\eqref{kvat-eq-1} we obtain the general
Euler\f Poisson equations with additional terms having different physical
interpretations~\cite{bormam2}.

As an interesting feature of system~\eqref{kvat-eq-1} we note that using
linear with respect to~$\lambda_i$ transformations we can reduce the
general form of the potential
\begin{equation}
\label{kvat-eq-2}
V=\sum_{i=0}^3 r_i\lambda_i
\end{equation}
to the form
\begin{equation}
\label{kvat-eq-3}
V=r_0\lambda_0.
\end{equation}
Indeed, the linear transformations of quaternion space~$\lambda_i$
(preserving commutation relations and quaternion norm) of the form
\begin{equation}
\label{lin-kvater}
\begin{aligned}
\wt\lambda_0&=R^{-1}(r_0\lambda_0+r_1\lambda_1+r_2\lambda_2+r_3\lambda_3),\\[10pt]
\wt\lambda_1&=R^{-1}(r_0\lambda_1-r_1\lambda_0-r_2\lambda_3+r_3\lambda_2),\\[10pt]
\wt\lambda_2&=R^{-1}(r_0\lambda_2+r_1\lambda_3-r_2\lambda_0-r_3\lambda_1),\\[10pt]
\wt\lambda_3&=R^{-1}(r_0\lambda_3-r_1\lambda_2+r_2\lambda_1-r_3\lambda_0),\\[10pt]
R^2\span = r_0^2+r_1^2+r_2^2+r_3^2
\end{aligned}
\end{equation}
reduce potential~\eqref{kvat-eq-2} to form~\eqref{kvat-eq-3}. The
existence of such linear transformation is the remarkable property of
quaternion variables and bracket~\eqref{eq-d2}, the analogous
transformation does not exists for the brackets of algebras~$e(3)$
and~$so(4)$.

In general dynamically unsymmetrical case~$a_1\ne a_2\ne a_3\ne a_1$
system~\eqref{kvat-eq-1} seems to be nonintegrable and none of two
necessary additional integrals exists. However, this was not proved, and
the proof by various reasons is not simple. Note that even the
application of the Kovalevskaya method\index{Method!Kovalevskaya}
for system~\eqref{kvat-eq-1} is not quite analogous to the classical
Euler\f Poisson problem. Generally speaking even in the Euler\f Poinsot
case the solution branches on the complex plane of time (with the
exponent~$1/2$).

At $a_1=a_2$ the linear integral always exists
\begin{equation}
\label{kvat-eq-4}
\begin{aligned}
F_1&=M_3(r_0^2+r_1^2+r_2^2+r_3^2)+N_3(r_1^2+r_2^2-r_0^2-r_3^2)+\\
{}&\quad + 2N_2(r_1r_0-r_3r_2)-2N_1(r_1r_2-r_0r_3),
\end{aligned}
\end{equation}
where~$N_i$ are projections of kinetic moment on the fixed axes. For
potential~\eqref{kvat-eq-3} this integral has the natural form
\begin{equation}
\label{kvat-eq-5}
F_1=M_3-N_3.
\end{equation}
It turns out that (linear) integral~\eqref{kvat-eq-5} corresponds to the
cyclic variable~$\vfi+\psi$~\cite{bormam2}. Rauss reduction with respect
to this cyclic variable results in Hamiltonian system on algebra~$e(3)$ on
zero surface of area integral~$(\bM,\bs\gamma)=0$ with the Hamiltonian
\begin{equation}
\label{kvat-eq-6}
H=\frac12(M_1^2+M_2^2+a_3M_3^2)+c(a_3-1)M_3+r_0\gamma_2+
\frac12\frac{c^2}{\gamma_3^2},
\end{equation}
where $c$ is a constant value of integral~\eqref{kvat-eq-5}.
Hamiltonian~\eqref{kvat-eq-6} corresponds to the addition of a gyrostatic
term linear with respect to~$\bM$ and singular term into the general
Euler\f Poisson equations. The physical meaning of singular term is
discussed in~\cite{bormam2}.

Integrable cases of system~\eqref{kvat-eq-1}
(equivalent to integrable cases of system~\eqref{kvat-eq-6}) are presented
in~\cite{bormam2}. Here we just present the generalization of Goryachev\f
Chaplygin case.

Hamiltonian and additional integral are
\begin{equation}
\label{kvat-eq-9}
\begin{aligned}
H&=\frac12(M_1^2+M_2^2+4M_3^2)+r_0\lambda_0,\\
F_2&=M_3(M_1^2+M_2^2)+r_0(M_2\lambda_1-M_1\lambda_2).
\end{aligned}
\end{equation}
Integral $F_2$ commute with integral~$F_1$. Under the reduction to
system~\eqref{kvat-eq-6} this case becomes a member of generalized
family~\eqref{gen5_1}.

\begin{rem}
Adding a constant gyrostatic moment along the axis of dynamical symmetry
in~\eqref{kvat-eq-9} we obtain an integrable case corresponding to the
generalized Sretensky case in Euler\f Poisson equations. Additional
integrals can be easily obtained with the help of the lifting procedure
described in~\cite{bormam2}.
\end{rem}

Note that the indicated ``Goryachev\f Chaplygin case'' for the quaternion
Euler\f Poisson equations is {\em the general integrable case}! Thus we
can use it for some algebraic constructions ($\bf L-A$\1pair constructions
and all that) described below.

\paragraph{Lax representation of systems on algebra $su(2,\,1)$.}
We start from some formal method of {\bf L\1A}~pair construction
based on existence of two
compatible Poisson brackets for the same system of Hamiltonian
equations on Lie algebra; in this case the system is called
bihamiltonian. The detailed exposition of this method and its
applications to various problems of rigid body dynamics are
presented in~\cite{BorisovMamaev, BolsBor}.

Let's consider space~$\cL$ of complex matrixes~$3\times 3$
with the basis
\eq[m9.1]{\alig{ {\bf M}_1 & =\left(\arr[c|c]{
\arr[cc]{0 & \frac 12 i \\ \frac 12 i & 0} & 0\\
\hline
0 & 0
}\right), & \qq
{\bf M}_2 & =\left(\arr[c|c]{
\arr[cc]{0 & \frac 12 \\ -\frac 12 & 0} & 0\\
\hline
0 & 0
}\right),\\
{\bf M}_3 & =\left(\arr[c|c]{
\arr[cc]{-\frac 12 i & 0 \\ 0 & \frac 12 i} & 0\\
\hline
0 & 0
}\right), & \qq
{\bf M}_4 & =\left(\arr[c|c]{
\arr[cc]{-\frac 16 i & 0 \\ 0 & -\frac 16 i} & 0\\
\hline
0 & \frac 13 i
}\right),\\
{\bf P}_1 & =\left(\arr[c|c]{
0 & \arr[c]{\frac 12 \\ 0}\\
\hline
-\frac 12 x\;\; 0 & 0
}\right), & \qq
{\bf P}_2 & =\left(\arr[c|c]{
0 & \arr[c]{-\frac 12 i \\ 0}\\
\hline
-\frac 12 x i\;\; 0 & 0
}\right),\\
{\bf P}_3 & =\left(\arr[c|c]{
0 & \arr[c]{ 0 \\ \frac 12 }\\
\hline
\frac 12 x i\;\; 0 & 0
}\right), & \qq
{\bf P}_4 & =\left(\arr[c|c]{
0 & \arr[c]{ 0 \\ -\frac 12 i }\\
\hline 0\;\; \frac 12 x i & 0 }\right). }
}

With respect to the standard matrix commutator~$[\cdot\,\,\cdot]$
such matrices generate a semi-simple algebra with Cartan
decomposition~$\cL=H+V $, where subalgebra~$H=su(2)\oplus su(1)$
is generated by matrices~${\bf M}_i$, and $V=\mC^2$ is generated
by matrices~${\bf P}_i$. Here~$x$~is a parameter determining some
bundle of algebras linearly dependent on~$x$. For~$x>0$ these
algebras are isomorphic to algebra~$su(3)$, for~$x<0$~they are
isomorphic to algebra~$su(2,\,1)$, at~$x=0$ they are isomorphic to
the semidirect sum $(so(2)\oplus su(1))\oplus_{s}\mC^2$.

Using the semi-simplicity we can identify algebra with coalgebra
by means of the inner product (Killing
form)
\eq[m9.2]{
g=-\Tr ({\bf X}\cdot {\bf Y}),\qq {\bf X},\,{\bf Y}\in \cL.
}

Let's denote coordinates in coalgebra
as~$m_1,\,m_2,\,m_3,\,m_4,\,p_1,\,p_2,\,p_3,\,p_4 $, then after
the identification we obtain a matrix (an element of algebra):
$$
 {\bf X}=\left(\arr[ccc]{
 -i(m_3+m_4) & im_1+m_2 & \frac 1x (p_1-ip_2)\\
 im_1-m_2 & i(m_3-m_4) & \frac 1x (p_3-ip_4)\\
-p_1-ip_2 & -p_3-ip_4 & 2i m_4
 }\right).
$$

The corresponding Lie\f Poisson bracket, more precisely, the
bundle of brackets linearly dependent on parameter~$x$ has the
following form for the coordinate functions of coalgebra
\eq[2.22*]{\alig{
\{m_i,\,m_j\} & =\eps_{ijk} m_k, & \q \{m_i,\,m_4\} & =0, \\
\{m_i,\,p_j\} & =\frac 12(\eps_{ijk}p_k-\dl_{ij}p_4), & \q
\{m_i,\,p_4\} & =\frac 12 p_i,
\qq i,\,j,\,k=1,\,2,\,3\\
\{p_i,\,p_j\} & =\frac 12 x(\eps_{ijk}m_k+3\eps_{ij3}m_4), & \q
\{p_i,\,p_4\} & =-\frac 12 x(m_i-3\dl_{i3}m_4).
}}

In correspondence with the general method developed
in~\cite{BorisovMamaev, BolsBor}, we shall construct~${\bf L}$
matrix, and its invariants define a commutative set for the whole
family of brackets $\{\;\}_\ta +\lm(\{\;\}_\lm +\{\;\}_a)$,
where~$ a\in V$ is a shift of the argument. The bracket $\{\}_\ta$
is different from~(\ref{2.22*}); in this bracket variables~$p_i$
are pairwise commuting. Let's assume~$x=1$ restricting ourselves
to the problem with a real dynamical sense. Then we can
present~${\bf L}$\1matrix as:
$$
 {\bf L}=({\bf h}\lm+{\bf v}+{\bf a}\lm^2),
$$
where
\eq[m9.3]{\alig{
{\bf h} & =\left(\arr[c|c]{
\arr[cc]{ {-}i(m_3{+}m_4) & im_1{+}m_2\\
im_1{-}m_2 & i(m_3{-}m_4)} & 0\\
\hline
0 & 2im_4
}\right)\\[10pt]
 {\bf v} & =\left(\arr[c|c]{
 0 & \arr[c]{ p_1-ip_2\\ p_3-ip_4}\\
 \hline
 {-}p_1{-}ip_2\;\; {-}p_3{-}ip_4 & 0
 }\right)\\[10pt]
 {\bf a} & =\left(\arr[c|c]{
 0 & \arr[c]{ a_1-ia_2\\ a_3-ia_4}\\
 \hline
 {-}a_1{-}ia_2\;\; {-}a_3{-}ia_4 & 0
 }\right). }}
Among invariants of matrix~${\bf L}$ under arbitrary shift
there is a linear on~$m_i$ function of the form
\eq[m9.4]{
F_1=m_4a^2+(\bs m,\,\bs \gam_a),
}
where~$a^2=\ds\sum_{i=1}^4 a_i^2,\;\bs m = (m_1,\,m_2,\,m_3)$,
and the components of vector~$\bs\gam_a$ are the functions
of~$a_i$ of the form
$$
 \bs\gam_a=(2(a_1a_3+a_2a_4),\,2(-a_2a_3+a_1a_4),\,a_3^2+a_4^2-a_1^2-a_2^2).
$$

Let's suppose~$a_1=a_2=0$. In this case integral~(\ref{m9.4})
has the form
\eq[m9.5]{
F_1=m_3+m_4.
}
The following square-law invariant of matrix~${\bf L}$, we will
choose as a Hamiltonian
\eq[m9.6]{
F_2=\Tr({\bf h}^2+2{\bf va})=m_1^2+m_2^2+m_3^2+3m_4^2+a_4p_4+a_3p_3.
}
It defines some (formal) integrable Hamiltonian system on the
family of brackets $\{\}_\ta +\lm(\{\}_\lm + \{\}_ a)$. We can
easily find matrix~${\bf A}$ for this system~\cite{BorisovMamaev}.

\paragraph{The reduced system and nonlinear Poisson structure.}
In order to proceed from the found formal system to the
generalization of Goryachev\f Chaplygin case~\cite{goriachev1}, we
shall make the reduction using linear integral~(\ref{m9.5})
directly in obtained~${\bf L}$ and~${\bf A}$ matrices. In the
general case this reduction is not Poisson
reduction~\cite{BorisovMamaev}. We shall make the following
substitution in matrix~${\bf L}$ (\ref{m9.2}) and in
Hamiltonian~(\ref{m9.6})
$$
  m_4=-m_3+c,\qq c=\const.
$$
We obtain~${\bf L}$ matrix and Hamiltonian of integrable system on
subalgebra~$m_1,\,m_2,\,m_3,\,p_1,\,p_2,\,p_3,\,p_4$ with
gyrostat, the gyrostatic moment being equal to~$c$:
\eq[m9.8]{
{\bf L}=\left(\arr[ccc]{
{-}i\lm c & (im_1{+}m_2)\lm & p_1{-}ip_2\\[10pt]
(im_1{-}m_2)\lm & i\lm(2m_3{-}c) & p_3{-}ip_4{+}(a_3{+}ia_4)\lm^2\\[10pt]
{-}p_1{-}ip_2 & {-}p_3{-}ip_4{-}(a_3{+}a_4)\lm^2 & {-}2i(m_3{-}c)\lm
}\right)
}
\eq[m9.8.5]{
H=m_1^2+m_2+4m_3^2-6m_3 c+2a_4p_4+2a_3p_3
}
\eq*{
{\bf A}=dH=\left(\arr[ccc]{
i(4m_3-3c) & -im_1-m_2 & 0\\
-im_1+m_2 & -i(4m_3-3c) & -a_3+ia_4\\
0 & a_3+ia_4 & 0
}\right).
}
Let's consider Poisson structure, defined by a bracket $\{\}_\ta$.
The corresponding Hamiltonian system with
Hamiltonian~(\ref{m9.8.5}) has another linear integral
\eqc[2.30]{
F_3=m_3-(\bs m,\,\bs\gam)\\
  \bs\gam=(2(p_2p_4+p_1p_3),\,2(p_2p_3-p_1p_4),\,p_3^2+p_4^2-p_1^2-p_2^2)),
}
using this integral we can carry out a usual procedure of order
reduction (such as Routh reduction or reduction on moment). We can
carry out the reduction in the simpler way using algebraic form,
if we choose as new variables the integrals of vector field $\bs
v=\{\cdot, F_3 \}$ of the form~\cite{BorisovMamaev}
\eqc[2.31]{
K_1=\frac{M_1p_1+M_2p_2}{\sqrt{p_1^2+p_2^2}},\q
K_2=\frac{M_2p_1-M_1p_2}{\sqrt{p_1^2+p_2^2}},\q
K_3=M_3,\\
s_1=p_3,\q s_2=p_4,\q s_3=\pm\sqrt{p_1^2+p_2^2},
}
having the nonlinear commutation. We describe the
transformation~(\ref{2.31}) in~\cite{brm}
\eqc[2.32]{
\{K_i,\,K_j\}=\eps_{ijk}K_k+\eps_{ij3}\frac{F_4}{s_3^2}\\
\{K_i,\,s_j\}=\eps_{ijk}s_k,\q \{s_i,\,s_j\}=0,
}
where $F_3 = (\bs K,\,\bs s) s_3 $ is a Casimir function of
structure~(\ref{2.32}).

At zero value of "the area integral", i.\,e. for $F_3=(\bs K,\,\bs
s) s_3=0$ bracket~(\ref {2.32}) coincides with a usual Lie\f
Poisson bracket, defined by algebra ${e(3)=so(3)\oplus_s\mR^3}$,
and Hamiltonian~(\ref{m9.8.5}) in variables~(\ref {2.31})
coincides with the Hamiltonian of the classical Goryachev\f
Chaplygin case~\cite{goriachev1}
\eq[2.33]{
H^*=\frac 12 H=\frac 12(K_1^2+K_2^2+4K_3^2)-a_4s_2-a_3s_1,\q
a_3,\,a_4=\const.
}

\paragraph{Goryachev case with a singular term.}
We can remove the nonlinearity of structure~(\ref{2.32}),
appearing for $F_3\ne 0$ on the fixed level $F_3=c$ using the
transformation
$$
\bs L=\bs K-c\frac{\bs s}{s_3}.
$$
After the transformation bracket~(\ref {2.32}) gets the form of
algebra~$e(3)$, and Hamiltonian~$H^*$ gets the form
\eq[2.34]{
H^*=\frac 12(L_1^2+L_2^2+4L_3^2)-a_4s_2-a_3s_1+3cL_3+\frac 12 \frac{c^2}{s_3^2}.
}
Hamiltonian~(\ref {2.34}) can be interpreted as some
generalization of the Goryachev\f Chaplygin case, at $(\bs L,\,\bs
s)=0$, with the additional terms, linear on $L_3 $ and
corresponding to the gyrostatic moment. The integrable
generalization with the gyrostatic moment only was described by
L.\,N.\,Sretensky in~\cite{sret}, the generalization with singular
potential only was presented by D.\,N.\,Goryachev
himself~\cite{goriachev1}, the general case, when it is possible to
add to the Hamiltonian both terms with arbitrary independent
coefficients, was described in paper~\cite{KomKuz}.

Thus, the presented {\bf L\1A}~pair is valid for the
generalizations of the Goryachev\f Chaplygin case. It is different
from the mysterious {\bf L\1A}~pair described in
paper~\cite{bobenko} which is obtained by deletion of a row and
column from the relevant pair of the Kovalevskaya case.

The authors thank A.\,V.\,Bolsinov, A.\,V.\,Tsyganov, V.\,V.\,Sokolov
for the valuable remarks and numerous discussions.


\newpage

\begin{thebibliography}{99}

\bibitem{bormam2}
    \author{A.\,V.\,Borisov, I.\,S.\,Mamaev}
    \title{Rigid body dynamics}
    \publisher{Izhevsk: RCD}
    \year{2001}
    \page{384}


\bibitem{KozlovMethods}
    \author{V.\,V.\,Kozlov}
    \title{Methods of qualitative analysis in rigid body dynamics}
    \publisher{Izhevsk: RCD}
    \year{2000}


\bibitem{bobenko}
    \author{A.\,I.\,Bobenko, V.\,B.\,Kuznetsov}
    \title{Lax representation and new formule for the Goryachev\f Chaplygin top}
    \journal{J. Phys. A}
    \volume{21}
    \year{1998}
    \page{1999--2000}

\bibitem{BolsBor}
    \author{A.\,V.\,Bolsinov, A.\,V.\,Borisov}
    \title{Lax representation and the compatible Poisson brackets on Lie algebras.}
    \journal{Math. notes (to be published)}


\bibitem{BorisovMamaev}
    \author{A.\,V.\,Borisov, I.\,S.\,Mamaev}
    \title{Poisson structures and Lie algebras in Hamiltonian mechanics}
    \publisher*{Izhevsk:~RCD publ.}
    \year{1999}
    \page{464}

\bibitem{brm}
    \author{A.\,V.\,Borisov, I.\,S.\,Mamaev}
    \title{Nonlinear Poisson brackets and isomorphisms in dynamics}
    \journal{Reg. \& Chaot. Dyn.}
    \volume{2}
    \year{1997}
    \no{3/4}
    \page{72--89}


\bibitem{goriachev1}
    \author{D.\,N.\,Goryachev}
    \title{New cases of rigid body motion around a fixed point}
    \journal{Warsaw univ. proceedings}
    \year{1915}
    \volume{3}
    \page{1--11}

\bibitem{KomKuz}
    \author{I.\,V.\,Komarov, V.\,B.\,Kuznetsov}
    \title{The generalized Goryachev\f Chaplygin gyrostst in
    quantum mechanics}
    \journal{Differential geometry, Lie groups and mechanics}
    \year{1987}
    \journal{Trans. of LOMI scientific seminar USSR Acad. of sciences}
    \volume{IX}
    \page{134--141}

\bibitem{sret}
    \author{L.\,N.\,Sretensky}
    \title{On some cases of integration of the gyrostat
    motion equations}
    \journal{Dokl. of USSR Acad. of Sciences. Mechanics}
    \volume{149}
    \no{2}
    \year{1963}
    \page{292--294}


\end{thebibliography}
\end{document}